\documentclass{ws-ijmpa}
\usepackage[super,compress]{cite}
\usepackage{graphicx}
\usepackage{multibib}
\usepackage{amsmath,amsfonts,amsthm,amssymb}
\usepackage[colorlinks=true,linktocpage=true,linkcolor=blue,citecolor=red,urlcolor=cyan]{hyperref}
\usepackage{lipsum}
\usepackage{wasysym}
\usepackage{cite}
\usepackage{mathtools}
\usepackage{xfrac}
\usepackage{afterpage}
\usepackage{slashed}
\usepackage{scalerel}
\usepackage[usenames,dvipsnames]{color}
\usepackage{subfigure}
\usepackage{bbold}
\usepackage{yfonts}
\usepackage{placeins}
\usepackage{bm} 
\usepackage{nicefrac}
\usepackage{soul}
\usepackage{mciteplus}
\usepackage{stackengine}
\usepackage{titlesec, blindtext, color}
\usepackage[perpage]{footmisc}
\renewcommand\L{\mathcal{L}}
\renewcommand\S{\mathcal{S}}
\newcommand\T{\mathcal{T}}
\newcommand\J{\mathcal{J}}
\usepackage{soul}
\newcommand{\olra}{\overleftrightarrow}

\usepackage{cancel}
\usepackage{slashed}
\begin{document}

%
\catchline{}{}{}{}{}
%

\title{On the non-uniqueness of the energy-momentum and spin currents}
\author{Rajeev Singh}
\address{Department of Physics, West University of Timisoara, Bulevardul Vasile P\^arvan 4, Timisoara 300223, Romania.\\
rajeev.singh@e-uvt.ro}
\maketitle

\begin{history}
\received{Day Month Year}
\revised{Day Month Year}
\end{history}
\begin{abstract}
The macroscopic energy-momentum and spin densities of relativistic spin hydrodynamics are obtained from the ensemble average of their respective microscopic definitions (quantum operators). These microscopic definitions suffer from ambiguities, meaning that one may obtain different forms of symmetric energy-momentum tensor and spin tensor through pseudogauge transformations (or, in other words, Belinfante improvement procedure). However, this ambiguity may be fixed if we obtain these currents using \emph{Noether's second theorem} instead of widely used Noether's first theorem. The second theorem fixes the super-potential determined by local symmetry, thereby selecting a unique physically consistent pseudogauge. In this article, we use Noether's second theorem to derive energy-momentum and spin currents without the need of pseudogauge transformations for free Dirac massive particles with spin one-half.
\end{abstract}
\ccode{PACS numbers:}
\tableofcontents
\section{Introduction}
\label{sec:intro}
The decomposition of the total angular momentum for a system of spin-half massive particles into orbital angular momentum and spin angular momentum is a long-standing issue that remains to be fully understood~\cite{Hehl:1976vr,Leader:2013jra,Freese:2021jqs}. In Einstein's general relativity (GR), the energy-momentum tensor (EMT) is symmetric due to its derivation with respect to the symmetric metric tensor \(g^{\mu\nu}\). However, when defining the EMT using Noether's first theorem from the Dirac Lagrangian, it comes out asymmetric, containing both symmetric and anti-symmetric components~\cite{Noether:1918zz,DeGroot:1980dk,Weinberg:1995mt}. This asymmetric EMT can be made symmetric using pseudogauge transformation (or in other words Belinfante improvement procedure)~\cite{BELINFANTE1939887,BELINFANTE1940449,Rosenfeld1940,Hehl:1976vr,Florkowski:2018fap,Speranza:2020ilk} resulting in the equivalence of Noether current and EMT from GR.

Indeed, it is always possible to redefine EMT and spin tensor through pseudogauge transformations, ensuring that the total charges—comprising energy, momentum, and angular momentum—remain unchanged~\cite{Hehl:1976vr}.
When focusing solely on these total charges, this redefinition ambiguity becomes inconsequential. However, in many physical scenarios, it is crucial to distinguish between the orbital and spin angular momentum of a system.

This raises questions about which definition to follow and the physical interpretation of the antisymmetric components of the EMT from the perspective of GR.
The important question we are having is: can a specific decomposition (or a specific pseudogauge) of total angular momentum for a given system provide a ``physically'' meaningful local distribution of energy, momentum, and spin?
While GR asserts that energy-momentum density is measurable through geometry, this matter is currently under intense debate, particularly when considering spin degrees of freedom~\cite{Florkowski:2018fap,Speranza:2020ilk}.


The other question comes to our mind is how to extract information from experiments to determine which pseudogauge describes the system. One possible approach involves recognizing that the expectation value of an observable in a state of local equilibrium generally depends on the pseudogauge. Moreover, in relativistic spin hydrodynamics, the values of the fields can vary depending on the pseudogauge used to decompose the total angular momentum. 
Understanding the decomposition of angular momentum into orbital and spin components is crucial for describing gauge fields. 
It is important to find a gauge-invariant way to decompose the total angular momentum. 

Such a way is, indeed, possible using Noether's second theorem. This theorem is applicable to local spacetime translations and local gauge transformations. Typically, the canonical EMT is derived using local coordinate transformations instead of global transformations. However, this approach is often applied inconsistently, treating the components of nonscalar fields as if they were a collection of scalar fields under these local translations~\cite{Itzykson:1980rh,Weinberg:1995mt}.
In this article, at first we mention and review the ambiguity present in the pseudogauge transformation and super-potential. Then using Noether's second theorem we re-derive the EMT expression and derive the spin tensor for massive spin-half fermions.

It is noteworthy to mention some of the issues and developments in this regard. The fundamental description of the spin and orbital angular momentum of light remains controversial~\cite{Bliokh:2014ara}. Recently, a gauge-invariant measure of photon spin, called the zilch current, has been studied in quantum kinetic theory and nuclear collisions~\cite{Chernodub:2018era,Huang:2020kik}. In hadron physics, angular momentum decomposition is vital for understanding quarks and gluons' contributions to the nucleon's spin~\cite{Leader:2013jra}. For works focusing on chiral physics and its connection to nuclear collisions see Refs.~\cite{Fukushima:2018osn,Fukushima:2020qta}.
\section{Relocalization of energy-momentum and spin currents}
\label{sec:pseudogauge_intro}
For a system of massive particles with spin, the total angular momentum tensor $({\J}^{\lambda, \mu\nu})$ is composed of orbital angular momentum tensor $({\L}^{\lambda, \mu\nu})$ and spin angular momentum tensor $({\S}^{\lambda, \mu\nu})$, namely~\cite{Itzykson:1980rh}
\begin{eqnarray}
{\J}^{\lambda, \mu\nu} = {\L}^{\lambda, \mu\nu} + {\S}^{\lambda, \mu\nu} = x^\mu {\T}^{\lambda\nu}-x^\nu {\T}^{\lambda\mu} + {\S}^{\lambda, \mu\nu}\,,
\label{eq:totalangM}
\end{eqnarray}
where $\T^{\mu\nu}$ is the EMT.

For a given Lagrangian density for any physical system and considering invariance of the action under spacetime translations and Lorentz transformations, to our disposal, we have two respective conservation laws using Noether's first theorem~\footnote{Noether's first theorem is most commonly known as Noether's theorem in the literature, however, this was the first theorem presented in Noether's original paper~\cite{Noether:1918zz}. Both first and second theorems of Noether imply conserved currents such as energy-momentum and total angular momentum.
The first theorem pertains to global symmetries such as global spacetime translations and global gauge transformations whereas the second theorem pertains to local symmetries such as local spacetime translations and local gauge transformations~\cite{Brading:2000hc,DeHaro:2021gdv}.}: 
\begin{itemize}
    \item conservation of EMT: 
    \begin{eqnarray}
        \partial_\mu
        {\T}^{\mu\nu} = 0\,,
        \label{eq:TangM}
    \end{eqnarray}
    \item and, conservation of total angular momentum:
    \begin{eqnarray}
        \partial_\lambda {\J}^{\lambda, \mu\nu} = 0\,.
        \label{eq:JangM}
    \end{eqnarray}
\end{itemize}
Total angular momentum conservation, using Eq.~\eqref{eq:totalangM}, gives
\begin{eqnarray}
\partial_\lambda {\J}^{\lambda, \mu\nu} = \partial_\lambda {\L}^{\lambda, \mu\nu} + \partial_\lambda {\S}^{\lambda, \mu\nu} &=& 0\,,\label{eq:DtotalangM1}\\
\partial_\lambda {\J}^{\lambda, \mu\nu}= {\T}^{\mu\nu} - {\T}^{\nu\mu} + \partial_\lambda {\S}^{\lambda, \mu\nu} &=& 0\,.
\label{eq:DtotalangM}
\end{eqnarray}
Let us do some analysis on the number of independent components of the above equation~\eqref{eq:DtotalangM1}. Since $\J$ is antisymmetric in $\mu$ and $\nu$, it has 24 independent components~\footnote{The general formula to find the number of independent components for a totally symmetric tensor of rank $r$ in $d$ spacetime dimensions is
\begin{eqnarray}
    \frac{(d+r-1)!}{r!\, (d-1)!}\,,
    \label{eq:symm_comp}
\end{eqnarray}
whereas the number of independent components for a totally antisymmetric tensor can be obtained using
\begin{eqnarray}
    \frac{d!}{r!\, (d-r)!}\,.
    \label{eq:antisymm_comp}
\end{eqnarray}}.
Moreover, terms in eq.~\eqref{eq:DtotalangM} has 6 independent components as the index $\lambda$ is summed up.

From eq.~\eqref{eq:DtotalangM}, we have
\begin{eqnarray}
\partial_\lambda {\S}^{\lambda, \mu\nu} + 2{\T}^{[\mu\nu]}_{\rm AS}&=& 0\,,
\label{eq:SangM}
\end{eqnarray}
where ${\T}^{\mu\nu}_{\rm AS}$ represents the antisymmetric part of the EMT with 6 independent components. The anti-symmetrization is notated by $A^{[\mu\nu]} = 1/2 (A^{\mu\nu} - A^{\nu\mu})$.
One may notice that Eq.~\eqref{eq:SangM} is another way of writing the total angular momentum conservation.

Equation~\eqref{eq:SangM} tells us that even though the total angular momentum is conserved, the spin angular momentum is not conserved independently due to the antisymmetric contributions from the EMT. One may also interpret it as the coupling between the orbital angular momentum and spin angular momentum or the source of spin current is the antisymmetric parts of ${\T}^{\mu\nu}$.

Let us go to a system of massive Dirac fermions of spin-half. For this system, the Lagrangian for the free fields (in the absence of any mean fields) is written as~\footnote{$\slashed{\partial}$ is $\slashed{\partial} = \gamma^\mu \, \partial_\mu$ and $\overleftrightarrow{\slashed{\partial}}  \equiv \overrightarrow{\slashed{\partial}} -\overleftarrow{\slashed{\partial}}$.}
\begin{eqnarray}
\mathcal{L}_{D}(x) = \frac{i}{2}\, \bar{\psi}(x)\, \overleftrightarrow{\slashed{\partial}}\,\psi(x)-m\, \bar{\psi}(x) \,\psi(x)\,,
\label{eq:Dirac_Lagrangian_0}
\end{eqnarray}
where $\psi$ and $\bar{\psi}\equiv \psi^{\dagger}\gamma^0$ is the Dirac field operator and its adjoint, respectively.
Using Noether's first theorem and equations of motion for the Dirac equation we can obtain the EMT as~\cite{Itzykson:1980rh}
\begin{align}
{\T}^{\mu\nu}_{\rm Can} = \frac{i}{2}\,\bar{\psi}\,\gamma^\mu\, \olra{\partial}^\nu\,\psi\,,
\label{eq:Can_e-m_tensor}
\end{align}
which is asymmetric. The subscript ``${\rm Can}$'' signifies the canonical procedure (using Noether's first theorem) to obtain the currents.

Equation~\eqref{eq:Can_e-m_tensor} can also be written in a form where one can explicitly identify the symmetric and antisymmetric parts
\begin{eqnarray}
    {\T}^{\mu\nu}_{\rm Can} &=& \frac{i}{4}\bar{\psi}\left(\gamma^{\mu}\overleftrightarrow{\partial}^\nu+ \gamma^{\nu}\overleftrightarrow{\partial}^\mu \right)\psi + \frac{i}{4}\bar{\psi}\left(\gamma^{\mu}\overleftrightarrow{\partial}^\nu - \gamma^{\nu}\overleftrightarrow{\partial}^\mu \right)\psi \,
\end{eqnarray}
where the first term is symmetric and second term is antisymmetric.
The spin tensor can be obtained in the form~\cite{Itzykson:1980rh}
\begin{eqnarray}
{\S}_{\rm Can}^{\lambda,\mu\nu} = \frac{i}{8}\,\bar{\psi}\Bigg\{\gamma^\lambda,\Big[\gamma^\mu,\gamma^\nu\Big]\Bigg\}\,\psi \,.
\label{eq:spin_canonical}
\end{eqnarray}
Note that, the spin tensor is completely antisymmetric in its indices having only 4 independent components.

Canonical EMT, eq.~\eqref{eq:Can_e-m_tensor}, seems to have some issues such as~\cite{CALLAN197042,Jackiw:2011vz,Blaschke:2016ohs,DiFrancesco:1997nk}
\begin{enumerate}
    \item For theories with gauge fields, it is not gauge invariant.
    \item If we would like spin to be conserved independent of total angular momentum conservation then ${\T}^{\mu\nu}_{\rm Can}$ needs to be completely symmetric. Moreover, in classical mechanics (neglecting spin for a moment) antisymmetric parts of ${\T}^{\mu\nu}_{\rm Can}$ prevent the total angular momentum conservation.
    \item It is not unique in linearized gravity.
    \item For theories which are manifestly scale invariant, ${\T}^{\mu\nu}_{\rm Can}$ is not traceless.
    \item Even though canonical global spin, ${\S}^{\mu\nu}_{\rm Can}$, satisfies $\rm{SO (3)}$ algebra~\cite{Dey:2023hft}, it is not a Lorentz tensor as it is not conserved due to the antisymmetric parts of ${\T}^{\mu\nu}_{\rm Can}$~\cite{Speranza:2020ilk}.
\end{enumerate}
    The conservation laws, Eqs.~\eqref{eq:TangM} and \eqref{eq:SangM}, are also fulfilled by the canonical currents. However, these forms of ${\T}^{\mu\nu}_{\rm Can}$ and ${\S}_{\rm Can}^{\lambda,\mu\nu}$ are not unique. These currents are determined only up to gradients, but by adding gradients these currents will be relocalized.

    The relocalized currents can be obtained through the pseudogauge transformations~\cite{BELINFANTE1940449,Hehl:1976vr,Speranza:2020ilk,Florkowski:2018fap}
    \begin{eqnarray}
    {\T}^{\mu\nu} &=& {\T}^{\mu\nu}_{\rm Can} + \frac{1}{2} \partial_\lambda ({\mathcal{X}}^{\lambda,\mu\nu}+{\mathcal{X}}^{\nu,\mu\lambda}+{\mathcal{X}}^{\mu,\nu\lambda})\,,\,\,\nonumber\\
    {\S}^{\lambda,\mu\nu} &=&{\S}^{\lambda,\mu\nu}_{\rm Can} - {\mathcal{X}}^{\lambda,\mu\nu} + \partial_\rho {\mathcal{Y}}^{\mu\nu,\lambda\rho}\,,
    \label{eq:pseudogauge_transformations}
    \label{lemma:LemmaII.1}
\end{eqnarray}
where the super-potentials ${\mathcal{X}}^{\lambda,\mu\nu}$ and ${\mathcal{Y}}^{\mu\nu,\lambda\rho}$ satisfy
\begin{eqnarray}
{\mathcal{X}}^{\lambda,\mu\nu} = -{\mathcal{X}}^{\lambda,\nu\mu}\,, \quad
{\mathcal{Y}}^{\mu\nu,\lambda\rho} = - {\mathcal{Y}}^{\nu\mu,\lambda\rho} = - {\mathcal{Y}}^{\mu\nu,\rho\lambda}\,.
\end{eqnarray}
The new currents still satisfy the conservation laws, \eqref{eq:TangM} and \eqref{eq:SangM}. The super-potential  ${\mathcal{X}}^{\lambda,\mu\nu}$ has 24 arbitrary functions whereas ${\mathcal{Y}}^{\mu\nu,\lambda\rho}$ has 36 arbitrary functions due to their symmetric properties.

The canonical currents and pseudogauge transformed currents represent different localization for energy, momentum, and spin. 
For gauge field theories in Minkowski spacetime the EMT needs to be gauge invariant due to its physical interpretation. But pseudogauge transformation does not yield a priori an EMT which is gauge invariant when applied to gauge theories. Additionally, it may not work straightforwardly for the physically interesting case such as when matter field and gauge field are minimally coupled. In the cases of electrodynamics and linearized Gauss-Bonnet gravity, the accepted physical, unique, gauge invariant, symmetric, conserved, and trace-free expressions are obtained from the Bessel-Hagen method~\cite{Bessel:1921cbh}. One might ask which one is preferred or correct (physical) localization out of so many possibilities.

    The total charges (i.e. total energy-momentum and total angular momentum)
    do not change under relocalization or pseudogauge transformation provided the super-potentials go to zero fast enough at space-like asymptotic infinity.

The most well known relocalized currents are obtained by assuming $${\mathcal{X}}^{\lambda,\mu\nu} = {\S}^{\lambda,\mu\nu}_{\rm Can}\,, \quad \text{and} \quad {\mathcal{Y}}^{\mu\nu,\lambda\rho}=0\,.$$ These will lead to Belinfante-Rosenfeld (BR) form of energy-momentum and spin tensors~\cite{BELINFANTE1939887,BELINFANTE1940449,Rosenfeld1940}, again using the equations of motion for the Dirac equation
    \begin{eqnarray}
     {\T}^{\mu\nu}_{\rm BR} = \frac{i}{4}\bar{\psi}\left(\gamma^{\mu}\overleftrightarrow{\partial}^\nu + \gamma^{\nu}\overleftrightarrow{\partial}^\mu \right)\psi\,, \quad
    {\S}^{\lambda,\mu\nu}_{\rm BR} = 0\,,
    \label{eq:BR_currents}
    \end{eqnarray}
    respectively.
    This relocalization reduces the total angular momentum to only orbital part by killing the Belinfante spin tensor.
    We note that since the canonical spin tensor is completely antisymmetric in nature, it basically kills the antisymmetric component of ${\T}^{\mu\nu}_{\rm Can}$.
    One may also obtain BR EMT using Euler-Lagrange equations as derived in Ref.~\cite{Weinberg:1995mt}.

Within the context of general relativity, the EMT is obtained by varying the Lagrangian with the metric that is equivalent with the BR EMT
\begin{eqnarray}
    {\T}^{\mu\nu}_{\rm H} = \frac{2}{\sqrt{-g}}\frac{\delta \L_D}{\delta g_{\mu\nu}}\,.
\end{eqnarray}
The motivation behind this (ad-hoc) procedure to obtain symmetric BR EMT using Noether theorem is to reproduce the Hilbert stress-energy tensor $\left({\T}^{\mu\nu}_{\rm H}\right)$ which is, by construction, conserved, symmetric, and gauge invariant and also seems to be the only viable EMT in classical GR.
\section{Ambiguity in super-potentials}
\label{sec:Noether_second_theorem}
Relativistic spin hydrodynamics is an extension of standard relativistic hydrodynamics that includes spin degrees of freedom.
Besides the conservation of energy and momentum, it includes spin degrees of freedom using total angular momentum conservation, Eq.~\eqref{eq:SangM}, as an extra hydrodynamic equation of motion. Relativistic hydrodynamics is a classical theory where energy-momentum density $(T^{\mu\nu})$ and spin density $(S^{\lambda,\mu\nu})$ are defined as the ensemble average of their respective quantum operators
\begin{eqnarray}
    T^{\mu\nu} = \langle \T^{\mu\nu} \rangle\,, \qquad S^{\lambda,\mu\nu} = \langle \S^{\lambda,\mu\nu} \rangle\,.
    \label{eq:densities}
\end{eqnarray}
To formulate such a formalism of relativistic spin hydrodynamics needs us to compute Eq.~\eqref{eq:densities} which in turn requires us to specify a pseudogauge. Pseudogauge transformation results to having symmetric EMT and then the spin tensor is conserved (when only local collisions are considered), however, as can be seen from Eq.~\eqref{eq:pseudogauge_transformations} that one can obtain different forms of EMT and spin tensor, hence, different formulations of relativistic spin hydrodynamics. Besides above mentioned BR form of EMT and spin tensor, two other forms of pseudogauges, within the context of spin hydrodynamics, are
\begin{itemize}
    \item de Groot--van Leeuwen--van Weert (GLW)~\cite{DeGroot:1980dk,Florkowski:2018fap} where  
    \begin{equation}{\mathcal{X}}^{\lambda,\mu\nu}=\frac{i}{4m}\bar{\psi}(\sigma^{\lambda\mu} \overleftrightarrow{\partial}^\nu-\sigma^{\lambda\nu}\overleftrightarrow{\partial}^\mu)\psi\,, \quad   {\mathcal{Y}}^{\mu\nu,\lambda\rho}=0\,,
    \end{equation}
    which yield
    \begin{eqnarray}
    {\T}^{\mu\nu}_{\rm GLW} &=& -\frac{1}{4m}\bar{\psi}\olra{\partial}^\mu\olra{\partial}^\nu\psi\,, \label{eq:GLW_currents}\\
    {\S}^{\lambda,\mu\nu}_{\rm GLW} &=& \bar{\psi}\left[\frac{\sigma^{\mu\nu}}{4}- \frac{1}{8m} \left(\gamma^{\mu}\olra{\partial}^\nu-\gamma^{\nu}\olra{\partial}^\mu \right)
    \right]\gamma^{\lambda} \psi + {\rm h.c}\,,\nonumber
    \end{eqnarray}
    with $\sigma^{\mu\nu} = (i/2)\left[\gamma^\mu , \gamma^\nu \right]$ being the Dirac spin operator.
        \item Hilgevoord--Wouthuysen (HW)~\cite{HILGEVOORD19631,HILGEVOORD19651002,Speranza:2020ilk} where
    \begin{eqnarray}
    {\mathcal{X}}^{\lambda,\mu\nu}&=& {M}^{[\mu\nu]\lambda}-g^{\lambda[\mu} {M}_\rho^{\ \nu]\rho}\,,\\
        \text{with} \quad {M}^{\lambda\mu\nu}&\equiv& \frac{i}{4m}\bar{\psi}\sigma^{\mu\nu}\olra{\partial}^\lambda \psi\,, \quad
        \text{and} \quad {\mathcal{Y}}^{\mu\nu,\lambda\rho}=-\frac{1}{8m}\bar{\psi}(\sigma^{\mu\nu}\sigma^{\lambda\rho}+\sigma^{\lambda\rho}\sigma^{\mu\nu})\psi\,,\nonumber
    \end{eqnarray}
    which yields
    \begin{eqnarray}
    {\T}^{\mu\nu}_{\rm HW} &=& {\T}^{\mu\nu}_{\text{Can}} +\frac{i}{2m} \left(\partial^{\nu}\bar{\psi} \sigma^{\mu\beta}\partial_\beta \psi +\partial_\alpha \bar{\psi} \sigma^{\alpha\mu}\partial^{\nu}\psi\right)- \frac{i}{4m}g^{\mu\nu} \partial_\lambda \left(\bar{\psi}\sigma^{\lambda\alpha}\overleftrightarrow{\partial}_\alpha \psi\right)\,,\nonumber\\
    {\S}_{\rm HW}^{\lambda,\mu\nu}&=& {\S}_{\rm Can}^{\lambda,\mu\nu}-\frac{1}{4m}\left( \bar{\psi} \sigma^{\mu\nu} \sigma^{\lambda\rho}\partial_\rho \psi + \partial_\rho \bar{\psi} \sigma^{\lambda\rho} \sigma^{\mu\nu}\psi\right)\,.\label{eq:HW_currents}
    \end{eqnarray}
\end{itemize}
In principle, one can have many forms of super-potential but the question still remains that which one describes the data or physical. This ambiguity is longstanding in both quantum field theory and gravity. At this point, it is just a choice of specifying a pseudogauge and develop the formalism of relativistic hydrodynamics with spin.
\section{Noether's second theorem and fixing of super-potential}
\label{subsec:noether_second_theorem}
In this section we would like to delve into Noether's second theorem and are particularly interested in its usefulness within the context of relativistic spin hydrodynamics~\cite{Florkowski:2017ruc,Florkowski:2018ahw,Hattori:2019lfp,Fukushima:2020ucl,Weickgenannt:2020aaf,Florkowski:2019qdp,Singh:2020rht,Florkowski:2021wvk,Hongo:2021ona,Ambrus:2022yzz,Singh:2022ltu,Singh:2022uyy,Gallegos:2022jow,Weickgenannt:2022jes,Becattini:2023ouz,Abboud:2025qtg}.
The latter is being currently investigated and developed in order to explain the spin polarization measurements~\cite{STAR:2023eck,STAR:2021beb,ALICE:2021pzu,ALICE:2019onw,STAR:2019erd,STAR:2017ckg}. These measurements have become an important probe to understand the properties of quark-gluon plasma. 

We would like to point out that Noether's second theorem can come to the rescue in resolving the ambiguity of super-potential. Using this theorem gives directly the symmetric EMT for free Dirac fields making the spin tensor conserved independently~\cite{Bessel:1921cbh,GamboaSaravi:2002vos,GamboaSaravi:2003aq,Montesinos:2006th,Freese:2021jqs} without specifying any super-potential. Hence, there is no need of pseudogauge transformations to make EMT symmetric.

Some remarks on the underlying principle of Noether's second theorem are as follows: once the Lagrangian and its variation of any physical system are fixed, Noether’s second theorem determines the super-potential. This applies not only to general relativity but to any theory with a local symmetry. In general relativity, the Noether current, i.e. gravitational pseudo-tensor, corresponding to local translation symmetry, is fixed by Noether’s second theorem through the choice of the action. In particular, the ability to add an arbitrary super-potential corresponds to the freedom to include boundary terms in the action without altering the equations of motion. Therefore, the super-potential is not arbitrary but is determined by Noether’s second theorem through the selection of boundary terms in the action.

The form of EMT for the free Dirac fields using Noether's second theorem is known, but the form of the spin tensor needs to be derived. Here we will provide the form of the spin tensor.
Let us briefly review the steps to obtain EMT from Noether's second theorem following Ref.~\cite{Freese:2021jqs}.

Consider the Lagrangian density as given in Eq.~\eqref{eq:Dirac_Lagrangian_0}.
We require the knowledge of the total variations $\Delta \psi$ etc., for further computations, which can be obtained requiring that
$\bar\psi \psi$ transforms as a scalar field,
$\bar\psi \gamma^\mu \psi$ transforms as a contravariant vector field, and the local translations, expressed as $\xi^\mu(x) = \epsilon\omega^{\mu}_{\phantom{\mu}\nu}x^\nu$ (where $\omega^{\mu}_{\phantom{\mu}\nu}$ is an antisymmetric tensor), replicate the established characteristics of spinors when subjected to infinitesimal Lorentz transformations.
Transformations which fulfil these conditions are
\begin{eqnarray}
  \label{eqn:variations:spinor}
    \Delta \psi
    &=&-
    \frac{1}{8}
    \partial_\alpha \xi_\beta
    \left[\gamma^\alpha,\gamma^\beta\right]
    \psi\,,
    \\
    \Delta \bar\psi
    &=&
    \frac{1}{8}
    \partial_\alpha \xi_\beta \,
    \bar\psi
    \left[\gamma^\alpha,\gamma^\beta\right]\,,
    \\
    \Delta \left(\partial_\mu\psi\right)
    &=&-
    \frac{1}{8} \partial_\alpha \xi_\beta
    \left[\gamma^\alpha,\gamma^\beta\right]
    \partial_\mu \psi
    -
    \partial_\mu \xi^\lambda \,
    \partial_\lambda \psi\,,
    \\
    \Delta \left(\partial_\mu \bar\psi\right)
    &=&
    \frac{1}{8}
    \partial_\alpha \xi_\beta\,
    \partial_\mu \bar\psi
    \left[\gamma^\alpha,\gamma^\beta\right]
    -
    \partial_\mu \xi^\lambda \,
    \partial_\lambda \bar\psi
    \,.\label{eqn:variationsq:spinor}
  \end{eqnarray}
Let us define a quantity $\mathcal{D}^{\mu\nu}$ as
\begin{eqnarray}
  \label{eqn:D}
  \mathcal{D}^{\mu\nu}\left[\psi_l\right]
  \equiv
  -
  \frac{\partial}{\partial\left(\partial_\mu \xi_\nu\right)}
  \left[
    \frac{\partial\mathcal{L}_D}{\partial \psi_l}
    \Delta \psi_l
    +
    \frac{\partial\mathcal{L}_D}{\partial (\partial_\mu\psi_l)}
    \Delta \left(\partial_\mu\psi_l\right)
    \right]\,,
\end{eqnarray}
where $l$ designates several fields.
Employing Noether's second theorem symmetries we obtain~\cite{Noether:1918zz}
  \begin{eqnarray}
    \mathcal{D}^{\mu\nu}[\psi]
    &=&
    - 
    \frac{\partial\mathcal{L}_{D}}{\partial\psi}
    \Sigma^{\mu\nu}
    \psi \,, \nonumber
    \\
    \mathcal{D}^{\mu\nu}[\bar\psi]
    &=&
    \bar\psi
\Sigma^{\mu\nu}
    \frac{\partial\mathcal{L}_{D}}{\partial\bar\psi} \,, \nonumber
    \\
    \mathcal{D}^{\mu\nu}[\partial\psi]
    &=&
    \frac{\partial\mathcal{L}_{D}}{\partial(\partial_\mu\psi)}
    \partial^\nu \psi
    -
    \frac{\partial\mathcal{L}_{D}}{\partial(\partial_\rho\psi)}
    \Sigma^{\mu\nu}
    \, \partial_\rho \psi \,, \nonumber
    \\
    \mathcal{D}^{\mu\nu}[\partial\bar\psi]
    &=&
    \partial^\nu \bar\psi
    \frac{\partial\mathcal{L}_{D}}{\partial(\partial_\mu\bar\psi)}
    +
    \left(\partial_\rho \bar\psi\right)
    \Sigma^{\mu\nu}
    \frac{\partial\mathcal{L}_{D}}{\partial(\partial_\rho\bar\psi)}
    \,,
    \label{eqn:D:spinor}
  \end{eqnarray}
where $\Sigma^{\mu\nu} = \left(i/4\right) \sigma^{\mu\nu}$.
Using the definition of EMT given as
\begin{eqnarray}
  \label{eqn:emt:general}
  \T^{\mu\nu}(x)
  =
  \sum_l
  \mathcal{D}^{\mu\nu}\left[\psi_l\right]
  -
  g^{\mu\nu} \mathcal{L}_{D}\,,
\end{eqnarray}
we receive
\begin{eqnarray}
    \T^{\mu\nu} (x) &=&
    \mathcal{D}^{\mu\nu}[\psi] + \mathcal{D}^{\mu\nu}[\bar\psi] + \mathcal{D}^{\mu\nu}[\partial\psi] + \mathcal{D}^{\mu\nu}[\partial\bar\psi]
    -
    g^{\mu\nu} \mathcal{L}_{D}\,,
      \nonumber\\
    &=&
    \frac{\partial \mathcal{L}_{D}}{\partial(\partial_\mu\psi)}
    (\partial^\nu \psi)
    +
    (\partial^\nu \bar\psi)
    \frac{\partial \mathcal{L}_{D}}{\partial(\partial_\mu\bar\psi)}
    -
    g^{\mu\nu} \mathcal{L}_{D}\nonumber\\
    && -
    \Bigg\{
      \frac{\partial \mathcal{L}_{D}}{\partial \psi}
      \Sigma^{\mu\nu}
      \psi
      -
      \bar\psi
      \Sigma^{\mu\nu}
      \frac{\partial \mathcal{L}_{D}}{\partial \bar\psi}+
      \frac{\partial \mathcal{L}_{D}}{\partial(\partial_\rho\psi)}
      \Sigma^{\mu\nu}
      (\partial_\rho\psi) -
      (\partial_\rho\bar\psi)
      \Sigma^{\mu\nu}
      \frac{\partial \mathcal{L}_{D}}{\partial(\partial_\rho\bar\psi)}
      \Bigg\}\,,
      \nonumber\\
    &=&
    \frac{i}{2} \bar\psi \gamma^\mu \overleftrightarrow{\partial}^\nu \psi
    -
    g^{\mu\nu}
    \bar\psi
    \left( \frac{i}{2}\overleftrightarrow{\slashed{\partial}} - m \right)
    \psi +
    \frac{1}{8}
    \Big\{
      \bar\psi
      [\gamma^\sigma, \sigma^{\mu\nu}]
      (\partial_\sigma\psi)
      -
      (\partial_\sigma\bar\psi)
      [\gamma^\sigma, \sigma^{\mu\nu}]
      \psi
      \Big\}
    \,,\nonumber\\
 &=&\frac{i}{4}\bar{\psi}\left(\gamma^{\mu}\overleftrightarrow{\partial}^\nu + \gamma^{\nu}\overleftrightarrow{\partial}^\mu \right)\psi\,.
  \label{eqn:emt:dirac}
\end{eqnarray}
It is remarkable that the EMT \eqref{eqn:emt:dirac} derived using Noether's second theorem matches with the BR form of EMT, see Eq.~\eqref{eq:BR_currents}, which was obtained using pseudogauge transformations.
It is noteworthy to comment that even though the form of EMT is known, however, it is derived without using the pseudogauge transformations and hence there is no ambiguity in the definition of super-potentials, see Eq~\eqref{lemma:LemmaII.1} and discussion thereafter.

Similarly, following the above procedure, we find that the spin tensor is actually vanishing.  
Inserting Eqs. \eqref{eqn:variations:spinor}–\eqref{eqn:variationsq:spinor} into definition \eqref{eqn:D}, substituting the Dirac Lagrangian $(\mathcal{L}_D=(i/2)\bar\psi \stackrel{\leftrightarrow}{\slashed{\partial}}\psi - m\bar\psi\psi)$, and using the identities
$(\partial \mathcal{L}_D/\partial(\partial_\lambda\psi)= (i/2)\bar\psi\gamma^\lambda)$ and $(\partial \mathcal{L}_D/\partial(\partial_\lambda\bar\psi)=-(i/2)\gamma^\lambda\psi)$, we get
\begin{eqnarray}
    \S^{\lambda,\mu\nu} (x)
    &=&\frac{1}{4}\Bigg[\frac{\partial \mathcal{L}_{D}}{\partial(\partial_\lambda \psi)}\gamma^\mu \gamma^\nu \psi -  \bar{\psi}\gamma^\mu \gamma^\nu \frac{\partial \mathcal{L}_{D}}{\partial(\partial_\lambda \bar{\psi})}\nonumber\\
    &&- \frac{\partial \mathcal{L}_{D}}{\partial(\partial_\lambda \psi)} \gamma^\mu \gamma^\nu \psi 
     + \bar{\psi} \gamma^\mu \gamma^\nu \frac{\partial \mathcal{L}_{D}}{\partial(\partial_\lambda \bar{\psi})} \Bigg]\,,
\end{eqnarray}
which yields
\begin{eqnarray}
    \S^{\lambda,\mu\nu}
    &=& 0\,.
    \label{eqn:spin:dirac_second}
\end{eqnarray}
The form of EMT, \eqref{eqn:emt:dirac}, has been favoured in a recent study~\cite{Das:2021aar} where it is shown that the energy-density quantum fluctuations, in subsystems of hot and relativistic spin-half massive particles, are very small for small system size which is not the case for other forms of EMT (GLW or HW). For large system sizes, energy-density quantum fluctuations obtained from Eq.~\eqref{eqn:emt:dirac} approaches faster towards thermodynamic limit.

In a recent study, Ref.~\cite{Dey:2023hft}, it is shown that the spin tensor, Eq.~\eqref{eqn:spin:dirac_second}, trivially satisfies SO(3) angular momentum algebra. It is also shown that the spin tensors (GLW or HW) obtained through pseudogauge transformations do not fulfill angular momentum algebra SO(3). It might be possible that any definition of EMT and spin tensor obtained through the use of super-potentials do not, in general, satisfy such algebra. However, this requires further investigations to have any solid conclusion.

It turns out that using Noether's second theorem we can naturally obtain EMT and spin tensor equivalent to classical GR currents.
\section{Summary}
\label{sec:summary}
In the context of relativistic spin hydrodynamics, macroscopic densities are obtained through the ensemble average of their respective quantum operators.
These microscopic quantum operators suffer from ambiguities as different forms of energy-momentum and spin tensors could be obtained through pseudogauge transformations. In this article, we have derived these quantum operators using Noether's second theorem, instead of Noether's first theorem, that pertains to local symmetries. 

We found that the derived EMT is symmetric in its indices and spin tensor is vanishing.
Noether's second theorem is not very widely known, however, using it, especially within the context of relativistic spin hydrodynamics, may remove the ambiguity of pseudogauge transformations and fix the super-potential. Though the EMT for the free fermions has been known, the spin tensor derivation using Noether's second theorem is mentioned for the first time in this article.
The method of derivation suggested in this paper could be useful to find suitable decompositions of momentum and spin for the comparison with other decompositions such as Ji~\cite{Ji:1996ek} and Jaffe-Manohar~\cite{Jaffe:1989jz}.
\section*{Acknowledgments}
We acknowledge important discussions with Adam Freese, C\'edric Lorc\'e, and Edward Witten.
We also thank the Institute for Nuclear Theory at the University of Washington for its kind hospitality and stimulating research environment. This research was supported in part by the INT's U.S. Department of Energy grant No. DE-FG02-00ER41132, and is also supported partly by a postdoctoral fellowship of West University of Timișoara.
\bibliographystyle{utphys}
\bibliography{references.bib}
\end{document}